\documentclass{elsart}
\input epsf.sty
\usepackage{natbib}
\begin{document}
\runauthor{Vangioni-Flam and Cass\'e}
\begin{frontmatter}
\title{Evolution of Lithium-Beryllium-Boron and Oxygen in the early Galaxy}

\author{Elisabeth Vangioni-Flam}
\author{ and Michel Cass\'e}
\address{Institut d'Astrophysique de Paris, CNRS, 
98 bis Boulevard Arago, 75014, Paris, France}
\address{Service d'Astrophysique, DAPNIA, CEA
Orme des Merisiers, 91191 Gif sur Yvette, CEDEX, France}
\begin{abstract}
Oxygen is a much better evolutionary index than iron to describe the history
 of Lithium-Beryllium-Boron (LiBeB) since it is the main producer of these
 light elements at least in the early Galaxy. The O-Fe relation is crucial
 to the determination of the exact physical process responsible for the LiBeB
 production. At low metallicity, if [O/Fe] vs [Fe/H] is flat,
 then the production mode is independent of the interstellar metallicity, BeB
 is proportional to oxygen, i.e. is of primary nature. If not, the production mode is function of
 the progressive enrichment in O of the interstellar medium, BeB varies rather as the square of O, i.e. 
 is of secondary nature. In the first case, fast nuclei enriched into He, C and O injected by
 supernovae and accelerated in surrounding superbubbles would explain the primary trend.
 In the second case, the main
 spallative agent would be the standard galactic cosmic rays. Calculated nucleosynthetic yields
 of massive stars, estimates of the energy cost of production of beryllium nuclei,
 and above all recent observations
 reported in this meeting seem to favor the primary mechanism, at least in the early Galaxy.
 
\end{abstract}
\end{frontmatter}

\section{Introduction}

Lithium-Beryllium-Boron take a special status in the general framework of nucleosynthesis.
These nuclei, are indeed of exceptional fragility and they are 
destroyed in stars above about 1 million degrees.
The only formation process available is spallation, i.e. 
fragmentation of medium light isotopes (CNO) leading to lighter 
species as ${^6}Li$, ${^7}Li$, ${^9}Be$, $^{10}B$ and $^{11}B$.
Lithium is special since it is involved in both the thermonuclear fusion
 (Big Bang and AGB stars and novae for $^{7}Li$), the neutrino spallation ($^{7}Li$ and 
 $^{11}B$ can be synthesized through break up $^{4}He$ and $^{12}C$ respectively)
 and nuclear non-thermal processes (both $^{6,7}Li$) Vangioni-Flam et al (1996).
The other light isotopes are pure nuclear spallation products 
The physical parameters  of the spallation mechanism are fourfold:
i) the production cross sections as a function of energy  are fairly well measured in the 
laboratory
ii) the source composition of the energetic component
ii) the associated (injection) energy spectrum.
iv) the target composition.
In the following, we describe the two possible spallative processes able to produce LiBeB
  and we confront them to observational constraints. We show how the 
 relation between oxygen and iron, in the halo
 phase is determining to discriminate between the two processes.

\section{Spallation processes and astrophysical sites}

The pioneering article of 
Meneguzzi, Audouze and Reeves (1971) offered the first 
quantitative explanation of the local abundances of 
LiBeB, at a time when only cumulated abundances in the solar system 
were available.
The standard Galactic Cosmic Rays (GCR), essentially composed of fast protons and alphas collide with 
CNO nuclei sitting in the interstellar medium to yield measured LiBeB abundances. 
But the observed isotopic ratios of Li and B were not reproduced. A stellar 
source of $^{7}Li$ was made necessary. For $^{11}B$  a complementary had hoc
spallative source of low energy was invoqued. Indeed, in this formulation evolution was ignored.

Introducing now the time parameter means to take into account the fact that the amount
 of CNO varies in the ISM  as well as the flux of cosmic rays (protons and alphas) $\phi$, presumably
  like the rate of supernovae, dN(SN)/dt, itself responsible for the increment of metallicity,
 we get the following equation: 

      $d(L/H)/dt = z(t) \langle \sigma \rangle \phi(t)$

 where z(t) is the evolving CNO fraction by number and $\langle \sigma \rangle$ is 
  the production cross section averaged over the energy spectrum. Since 
 z is proportional to N(SN), integrating we get, assuming a constant spectral shape:

  $d(L/H)/dt \alpha N(SN) dN(SN)/dt$ or $L/H \alpha z^2$
 
 Then the abundance of a given light element
 increases like the square of the CNO abundance (or as a good approximation to O).
  Thus according to the 
classical tradition in galactic evolution, the production of LiBeB by 
the GCR is called "secondary".

In the nineties, measurements of Be/H and B/H from KECK and HST, together 
with [Fe/H] (for a report of observations see IAU Symposium 198, 2000) in very low metallicity
 halo stars came to set 
strong constraints on the origin and evolution of LiBeB isotopes.
 The evolution of BeB was suddenly uncovered  over about 10 Gyr, taking [Fe/H] 
as an evolutionary index.
The linearity between Be, B and iron came as a surprise since a quadratic 
relation was expected from the standard GCR mechanism. It was a strong
 indication that the standard GCRs are not the main producers of LiBeB in
 the early 
Galaxy. A new mechanism of primary nature was required to
 reproduce these observations: it has been proposed that low energy CO nuclei ( a few tens MeV/n)
 produced and accelerated by massive stars (WR and SN II in superbubbles, i.e. cavities in
 the interstellar medium excavated by the winds and explosion of massive stars) fragment
  on H and He
 at rest in the ISM. This low energy component (LEC) has the advantage of
 coproducing Be and B in good agreement with the ratio observed in Pop II
 stars (see figure 1) (Vangioni-Flam et al 2000, Ramaty et al 2000).
 The corresponding equation is then:

  $d(L/H)/dt = n(H, He) \langle \sigma \rangle \phi(CO)(t)$

 where the first term is the concentration of H and He in the ISM, 
approximatelly constant,  $\langle \sigma \rangle$ is the cross section averaged over the 
equilibrium (intestellar) spectrum, and  $\phi(CO)(t)$
  the flux of C and O 
nuclei, assumed to be proportional to the supernova rate.
Owing to the constancy of the abundance of the target nuclei, the 
integration leads to a strict proportionality between the 
cumulated Be, B abundances and the metallicity. The LEC process is 
naturally called "primary".

\begin{figure}[h]
	\centering
\vspace{8.cm} 
\includegraphics{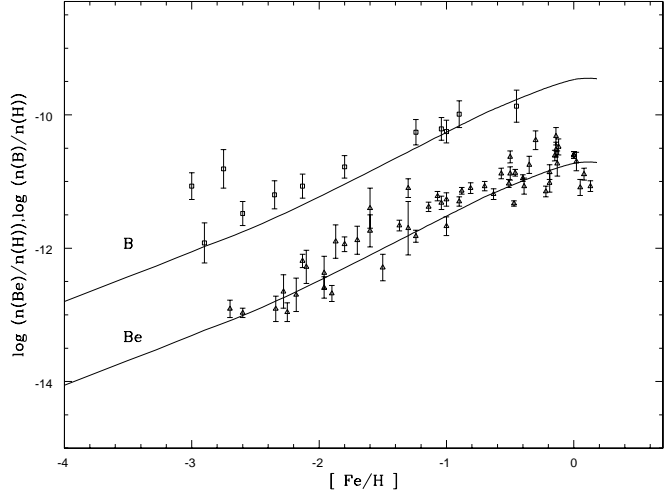}
\caption{Beryllium and Boron evolution vs [Fe/H]. The halo evolution
  is dominated by the LEC component (primary) emanating from supernovae. 
 The contribution of the standard GCR component (secondary) is unsignificant below
 [Fe/H] $<$ -1 if O is proportional to Fe, if not, i.e. [O/Fe] $\alpha$ [Fe/H], the
 transition between primary and secondary processes could occur around
 [Fe/H] = -2, Vangioni-Flam et al 1998, Fields and Olive 1999.}  
\label{fig1}
\end{figure}

Note that the two processes have in common the accelerating agents, namely the 
supernovae. In our opinion, they are not mutually exclusive but 
rather successive, the primary process being active in the halo phase and the secondary in the
 disk. The main differences are: i) The source composition of LEC, reflecting that of SN ejecta,  is 
enriched into C, O and He. The richness in alpha particles, compared to the standard GCR ones,
  allows in particular a generous formation of $^{6}Li$ (Vangioni-Flam et al 1999) which can be
invoked to explain the abundance of this  isotope in halo stars. 
ii) The form of the energy spectrum: LEC could have a different 
spectrum than GCR, since the conditions of acceleration in 
Superbubbles are different from those prevailing in the interstellar 
medium (Bykov 2000). The spectrum is predicted to 
be enriched in low energy particles. The mean energy of LEC, 
averaged on time being estimated to about 100 MeV/n (Bykov, private communication),
 whereas that of GCR is of the order of 1GeV/n.
iii) Above all, they differ by the order of the production 
process, primary for LEC and secondary for standard GCR.
Today, since data are available down to [Fe/H] = -3,
 the evolutionary models have not only to explain the cumulative 
abundances measured in the solar system but also the evolutionary trends.

The term "metallicity", up to now has been ambiguously 
 defined. In fact, all the above argumentation
assumes that CNO/Fe is constant at [Fe/H] less than -1.
Since oxygen is the main progenitor of BeB, the
 apparent linear relation between BeB and Fe could be misleading if O was
 not strictly proportional to Fe. Thus the pure primary origin
 of BeB in the early Galaxy could be questionned.

If O/Fe is constant, the GCR process is 
clearly negligible due to the paucity of the ISM in CNO, in the halo phase. The LEC process is 
obviously predominent, since it is free from the ISM metallicity, 
relying on freshly synthesized He, C, and O. Progressively, 
following the general enrichment of the ISM, GCR gain importance.
Neutrino spallation plays its marginal role, increasing the abundance 
of $^{11}B$.

Recent observations of Israelian et al  (1998) and Boesgaard et al (1999) (IB) 
showing a neat slope in the O-Fe plot have seeded a trouble. If 
this trend is verified, the whole interpretation has to be modified, giving a larger role
 to the secondary process in the halo phase (Fields ans Olive 1999). Nevertheless, even in this case, 
 according the general consensus, the primary component is at least required
 in the very early Galaxy. However 
these observations are considered as controversial, and the whole 
session is centered on this point.
What is the right O-Fe correlation? This point is crucial to translate 
the (Be, B) - Fe relations into (B, Be) - O ones.
In the IB observations, [O/Fe]=-0.35 [Fe/H], and consequently log (Be/H)
 is  proportional to 1.55 [O/H]. Fields and 
Olive have used this relation to rehabilitate the classical standard 
 GCR as the progenitor of LiBeB. But, beyond the observational questioning, this scenario
 meets with theoretical difficulties. The 
energy cost to produce a single Be nucleus is unfavorable but not prohibitive since this
 cost is plagued by large uncertainties (Fields et al 2001).
Another difficulty is that the stellar supernova yields integrated in 
a galactic evolutionary model cannot fit the new O/Fe data (\cite{EVF2}, F. Matteucci, this meeting).
Moreover, the observational [$\alpha$,/Fe] vs [Fe/H] where $\alpha$ = Mg, Si, Ca, Ti show a plateau
 from [Fe/H] = -4 to -1. It would be surprising that oxygen does not follow the Mg, Si and Ca trends 
 since all these nuclei are produced by the same massive stars.
If the trend expressed in this meeting, namely [O/Fe] is approximatively constant or slightly varying
 in the halo phase is confirmed, then the theoretical situation is clarified and the primary component
 made necessary in the early Galaxy, until [Fe/H]= -1.

\section{Conclusion}

 The synthesis of LiBeB in the halo proceeds through nuclear spallation, essentially by
 the break up of oxygen. The observed relation between BeB and Fe has to be translated through
 the O - Fe relation into a BeB - O one which is representative of the relevant physical production
 process. And consequently the observational O - Fe relation is determining. 
(except if one succeeds to measure Be, B and O simultaneoulsy in 
stars (A.M. Boesgaard and K. Cunha, this 
meeting). Anyway the primary component is required in the halo phase, then afterwards the secondary
 process takes over; the question is therefore
 at which metallicity the transition occurs, this depends on the behaviour of [O/Fe]. 
  To test in details the scenarios, the ideal would be to get the evaluation of
 O, Fe, Mg Be, B, Li abundances in the same stars. Impressive progress are waited
 from the VLT. 
Gamma ray line astronomy, through the european INTEGRAL 
satellite, to be lauched in 2002, will help to constrain the energy 
spectrum and intensity of LEC.


\begin{thebibliography}{9}
\bibitem A.M. Boesgaard, J.R. King, C.P. Deliyannis and S.S. Vogt,
 Astron. J. {\bf 117} (1999) 492.
\bibitem A. Bykov, Space Science Rev. {\bf 100} (2000) 1.
\bibitem B.D. Fields and K.A. Olive, Astrophys. J. {\bf 516} (1999) 797.
\bibitem B.D. Fields, K.A. Olive, M. Cass\'e and E. Vangioni-Flam, Astron. Astrophys. 
 (2001) in press.
\bibitem IAU Symposium 198, "The light elements and their Evolution", Eds L. da Silva,
 R. de Medeiros, M. Spite, ASP Conf. Series (2000) in press.
\bibitem G. Israelian, R. F. Garcia-Lopez, R. Rebolo, Astrophys.J. 
 {\bf 507}(1998) 357.
\bibitem M. Meneguzzi, J. Audouze, H. Reeves, Astron.Astrophys. {\bf 15}(1971)337.
\bibitem R. Ramaty, S.T. Scully, R.E. Lingenfelter and B. Kozlovsky, Astrophys. J.,
 {\bf 534} (2000) 747.
\bibitem E. Vangioni-Flam, R. Ramaty, K.A. Olive and M. Cass\'e, Astron.Astrophys. {\bf337}
 (1998) 714.
\bibitem E. Vangioni-Flam, M. Cass\'e, B.D. Fields, K.A. Olive, Astrophys. J. 
{\bf 468} (1996) 199.
\bibitem E. Vangioni-Flam, M. Cass\'e, J. Audouze, Phys. Rep. {\bf 333-334}
 (2000) 365.
\bibitem E. Vangioni-Flam et al, New Astron. {\bf 4} (1999) 245.
\end{thebibliography}
\end{document}